%% file: main.tex
\documentclass[10pt,conference]{IEEEtran}
\usepackage[utf8]{inputenc} 
\usepackage[T1]{fontenc}
\usepackage{url}

\usepackage{breakurl}
\usepackage{ifthen}
\usepackage{graphicx}
\usepackage[T1]{fontenc}

\usepackage{enumitem}
\usepackage[cmex10]{amsmath}
\usepackage{amssymb}
\usepackage{multicol}

\usepackage{cite}
\usepackage{amsthm}
\usepackage{textcomp}
\usepackage{siunitx}
\usepackage{color}
\usepackage{cases}
\usepackage[font={small}]{caption}
\usepackage{epstopdf}
\usepackage{graphicx}
\usepackage{caption}
\usepackage{verbatim} 
\usepackage{bbm}
\usepackage{mathtools}

\newtheorem{theorem}{Theorem}
\newtheorem{lemma}{Lemma}

\newtheorem{remark}{Remark}

\theoremstyle{definition}

\theoremstyle{remark}

\newtheorem*{theorem*}{Theorem}

\usepackage[ruled,vlined]{algorithm2e}
\SetAlFnt{\footnotesize} 
\setlist[description]{style=multiline}

\begin{document}
\sloppy
\title{Carbon-Neutralized Joint User Association and Base Station Switching for Green Cellular Networks}
    
\author{
\IEEEauthorblockN{Chien-Sheng Yang, Carlson Lin and I-Kang Fu}
\IEEEauthorblockA{MediaTek Inc.
\\
Email: \{chien-sheng.yang, carlson.lin, ik.fu\}@mediatek.com}
}
\maketitle

\begin{abstract}
\input{0-abstract}
\end{abstract}

\section{Introduction}
\input{1-intro.tex}

\section{System Model} \label{sec:sys}
\input{2-system.tex}
\section{Problem Statement}\label{sec:MINLP}
\input{3-MINLP.tex}

\section{The Proposed Carbon-Aware Scheme}\label{sec:MILP}
\input{4-MILP.tex}

\section{Numerical Analysis}
\input{5-analysis.tex}
\section{Conclusion}
\input{6-conclusion.tex}

\bibliographystyle{ieeetr}
\bibliography{references}
\appendices
\input{Appendix.tex}

\end{document}

%% file: 0-abstract.tex
Mitigating climate change and its impacts is one of the sustainable development goals (SDGs) required by United Nations for an urgent action. Increasing carbon emissions due to human activities is the root cause to climate change. Telecommunication networks that provide service connectivity to mobile users contribute great amount of carbon emissions by consuming lots of non-renewable energy sources. Beyond the improvement on energy efficiency, to reduce the carbon footprint, telecom operators are increasing their adoption of renewable energy (e.g., wind power). The high variability of renewable energy in time and location; however, creates difficulties for operators when utilizing renewables for the reduction of carbon emissions. In this paper, we consider a heterogeneous network consisted of one macro base station (MBS) and multiple small base stations (SBSs) where each base station (BS) is powered by both of renewable and non-renewable energy. Different from the prior works that target on the total power consumption, we propose a novel scheme to minimize the carbon footprint of networks by dynamically switching the ON/OFF modes of SBSs and adjusting the association between users and BSs to access renewables as much as possible. Our numerical analysis shows that the proposed scheme significantly reduces up to $86\%$ of the non-renewable energy consumption compared to two representative baselines.

%% file: 1-intro.tex

Climate change due to escalating carbon emissions has emerged as one of the most critical environmental challenges to Earth. To reverse the effects of climate change that has posed a serious impact to our world (e.g., more frequent and severe droughts), the Intergovernmental Panel on Climate Change (IPCC) has set the goal to reach net-zero carbon emissions by mid-century~\cite{2022ipcc}. To achieve carbon neutrality, one simple approach is to compensate carbon emissions by acquiring carbon offsets, e.g., UN Carbon Offset Platform that has been widely adopted globally for trading carbon credits. The limitations of such mechanisms~\cite{2022cssn} has pushed the development of low-carbon technologies that reduce carbon emissions at the beginning rather than offset them later. 
 
Recent advancements in cellular technologies (e.g., 5G) that enable a wide range of applications (e.g., virtual reality (VR), augmented reality (AR), Industry 4.0, smart cities) has led to an explosive growth of service demands in networks. Following this trend, ICT sector is expected to account for $20\%$ of the global energy consumption by 2040~\cite{GreenG}. Thus, the carbon emissions caused by the surge of energy consumption in cellular networks require a significant enhancement on energy efficiency to mitigate the resulting negative impacts to our world. To ensure that the next generation cellular network (i.e., 6G) can be developed toward the goal of carbon neutrality, the new initiatives have also been launched, e.g., the Green G Working Group in Next G Alliance~\cite{GreenG}. 




Base station (BS) is the most energy-intensive part of a cellular network~\cite{6056691}. Such energy intensity is further amplified in a heterogeneous network where an increasing number of small base stations (SBSs) are deployed to enhance capacity of a macro base station (MBS). Due to the change of traffic pattern over time and location, some of under-utilized SBSs consume unnecessary energy. Given the fact that a user in the overlapping coverage areas of some SBSs can be served by any one of them, BS ON-OFF switching has emerged as one promising approach to improve energy efficiency of low-load SBSs by dynamically adjusting their ON/OFF modes~\cite{wu2015energy}, which is being studied in 3GPP (e.g., \cite{3gpp.28.310}). Another key approach for telecom operators to reduce their carbon footprint is utilizing more renewable energy that does not release carbon dioxide when producing electricity (e.g., AT$\&$T~\cite{ATT}).

The current BS ON-OFF switching design mainly focus on minimizing the total energy consumption (i.e., the sum of non-renewable and renewable energy consumption), which makes the ideas of saving energy and using more renewables two separate approaches. Meanwhile, the availability of renewable power is highly variable in time and location~\cite{callaway2018location}. Furthermore, the operation of BS ON-OFF switching also involves user association (i.e., users served by one BS that is switched to OFF mode should be associated with other BS in ON mode), which needs to take the users' random behavior into account. Therefore, to guarantee the reduction of carbon emissions in cellular networks, it is desired to design an effective scheme that jointly adjusts the ON/OFF modes of SBSs and the association between users and BSs while utilizing highly variable renewable power as much as possible.

Different from the prior works that target on the total power consumption (e.g., \cite{7997144,8644146}), we study the problem of joint user association and BS ON-OFF switching with particular focus on the carbon footprint of non-renewable power consumption. We consider a heterogeneous network consisted of one MBS and multiple SBSs where each BS is powered by both of renewables and non-renewables. While the MBS is assumed to be always in ON mode, each SBS can be switched to ON or OFF mode. Each user can be served by a BS in ON mode if and only if the user is located at its service coverage. To capture the behavior that an idle or low-load BS still consumes power, we model the power consumption of each ON-mode BS out of a linear function with a static power component (e.g., cooling). With the objective of minimizing the total non-renewable power consumption of network, we formulate a mixed integer nonlinear programming (MINLP) that jointly decides the mode of each SBS and the association between BSs and users.

The formulated MINLP problem with a nonlinear objective function of the non-renewable power consumption; nevertheless, is an intractable combinatorial problem~\cite{lee2011mixed}. Through a derived linear upper bound for approximating the objective function, we propose a carbon-aware scheme by formulating a mixed integer linear programming (MILP), which can be solved efficiently in practice. Compared with two baseline schemes, our numerical analysis shows that the proposed carbon-aware scheme reduces up to $86\%$ of the non-renewable power consumption, which can potentially save billions of US dollars annually for the cost of carbon tax. 

\noindent{\bf Related Works:} We provide a literature review that covers the works of carbon-aware network and BS ON-OFF switching.

The development of carbon-aware network has received much attention recently to mitigate the global warming issue. One of main goals is to do resource allocation by adapting to the variability of renewable energy (see e.g., \cite{9770383,yang2022carbon,8443373}). To reduce carbon footprint of cloud networks,~\cite{9770383} introduced a Carbon-Intelligent Compute Management by shifting flexible compute workloads based on the future carbon-related information. Via the drift-plus-penalty methodology in Lyapunov optimization, \cite{yang2022carbon} proposed a carbon-intensity based scheduling policy that dynamically schedules computation tasks over cloud networks. In heterogeneous cellular networks powered by both of renewables and non-renewables,~\cite{8443373} presented a user association mechanism that efficiently utilizes the available renewable power.

The idea of BS ON-OFF switching that aims at dynamically turning off some under-utilized BSs during low traffic periods, has been widely studied in recent years to improve energy efficiency of cellular networks. Several works investigated BS ON-OFF switching over different system models (e.g., \cite{ 8885821, 9500442, 6907934, 7997144, 8644146}). Considering the inter-tier interference between the small cells, BS ON-OFF switching is leveraged in cloud radio access networks \cite{8885821}. Through a long short-term memory (LSTM) method to predict traffic distribution, a switching scheme proposed in \cite{9500442} can reduce the energy consumption of networks while guaranteeing the quality of experience of users. Powered by renewable power sources,~\cite{6907934} proposed a dynamic programming-based BS ON-OFF switching to minimize a sum of non-renewable power consumption and the quality of service. Furthermore, \cite{7997144,8644146} consider the problem of joint user association and BS ON-OFF switching to minimize the total power consumption of heterogeneous cellular networks. To distinguish from the prior works of joint user association and BS ON-OFF switching which target on improving the efficiency of total power consumption, our proposed scheme focus on minimizing non-renewable power consumption, which accounts for the majority of carbon footprint from cellular networks.

%% file: 2-system.tex
\begin{figure*}[t]
    \centering
    \includegraphics[width = 0.6\linewidth]{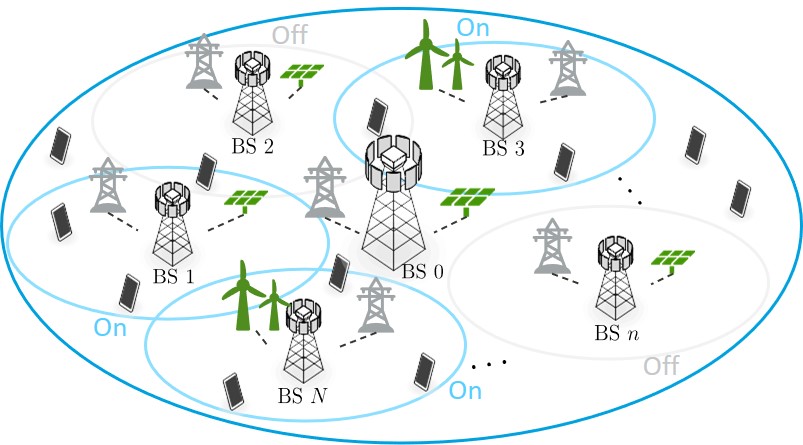}
    \caption{The illustration of considered network including one MBS and $N$ SBSs. Each BS is powered by both of non-renewables and renewables. The MBS is always in ON mode while each SBS can be switched to ON or OFF mode.}  
    \label{fig:sys}
\end{figure*}
We consider a heterogeneous cellular network, composed of a macro base station (MBS) and multiple small base stations (SBSs), provides service to users. All SBSs are deployed in the service coverage of the MBS to enhance the network capacity. In particular, each of base stations (BSs) is powered by both of non-renewable energy (e.g., fossil) and renewable energy (e.g., wind). The MBS is always turned on to provide main service coverage whereas each of SBSs can be switched to ON or OFF mode according to the location of users and the amount of available renewable power.

In the network, as shown in Fig.~\ref{fig:sys}, there are one MBS and $N$ SBSs serving $M$ users within a geographical area $\mathcal{L} \subseteq \mathbb{R}^2$. We denote by $\mathcal{N}= \{0,1,\ldots,N\}$ the set of all base stations where the MBS is represented by the index of $0$ for the simplicity. We denote by $\mathcal{M} = \{1,2,\ldots, M\}$ the set of users. With the coverage radius $r_n$, each BS $n \in \mathcal{N}$ is located at the coordinate denoted by $(x^{\text{BS}}_n, y^{\text{BS}}_n) \in \mathcal{L}$. Moreover, each user $m \in \mathcal{M}$ is located at the coordinate denoted by $(x^{\text{user}}_m, y^{\text{user}}_m) \in \mathcal{L}$. We denote by $d_{m,n}$ the distance between user $m$ and BS $n$, i.e., $d_{m,n} = \sqrt{(x^{\text{user}}_m-x^{\text{BS}}_n)^2 + (y^{\text{user}}_m - y^{\text{BS}}_n)^2}$. In the following, we describe the operations of base station ON-OFF switching and user association.

{\bf Base Station ON-OFF Switching:} In our model, BS $0$ (i.e., MBS) is always in ON mode. Each BS $n \in \mathcal{N}\backslash \{0\}$ (i.e., SBSs) can be switched to ON or OFF mode, where BS in OFF mode cannot provide service to any users. Specifically, the mode of BS $n$ is indicated by $a_n$ as follows:
\begin{align}
    a_{n} = 
\begin{cases}
    1, \ \text{if BS $n$ is ON;}\\
    0, \ \text{if BS $n$ is OFF}.
\end{cases}
\end{align}
In ON mode, each BS $n$ can serve user $m$ if user $m$ is within the coverage of BS $n$, i.e., $d_{m,n}\leq r_n$. We denote by $\mathcal{S}^{\text{BS}}_n$ the set of users that are able to be served by BS $n$, i.e.,
\begin{align}
    \mathcal{S}^{\text{BS}}_n = \{m: m \in \mathcal{M}, d_{m,n}\leq r_n\}.
\end{align}
Similarly, we denote by $\mathcal{S}^{\text{user}}_m$ the set of BSs that are able to serve user $m$, i.e.,
\begin{align}
    \mathcal{S}^{\text{user}}_m = \{n: n \in \mathcal{N}, d_{m,n}\leq r_n\}.
\end{align}
In particular, we denote by $B_n$ the maximum number of users that can be served by BS $n$ in ON mode. 

{\bf User Association:} After the operation of BS ON-OFF switching, each user $m$ still has to be served by one of BSs, i.e., users served by the BS that is switched to OFF mode have to be associated with other BS in ON mode. Concretely, we denote by $w_{m,n}$ the indicator whether BS $n$ serves user $m$ as follows:
\begin{align}
    w_{m,n} = 
\begin{cases}
    1, \ \text{if user $m$ is served by BS $n$;}\\
    0, \ \text{otherwise}.
\end{cases}
\end{align}
To guarantee that each user is served by exactly one BS, the followings have to be satisfied:
\begin{align}
    \sum_{n \in \mathcal{S}^{\text{user}}_m}w_{m,n} = 1,&\ \forall m  \in \mathcal{M};  \label{eq:condition1}\\
    w_{m,n} \leq a_n,& \ \forall m  \in \mathcal{M}, \ \forall n  \in \mathcal{N}. \label{eq:condition2} 
\end{align}
where \eqref{eq:condition1} indicates that user $m$ is served by one of BSs and \eqref{eq:condition2} ensures that each BS can serve users only if the BS is in ON mode. Moreover, the number of users served by each BS $n$ has to be smaller than $B_n$, i.e., 
\begin{align}
    \sum_{m \in \mathcal{S}^{\text{BS}}_n}w_{m,n} \leq B_n,&\ \forall n  \in \mathcal{N}.  \label{eq:condition3}
\end{align}  

{\bf Power Consumption Model:} Based on~\cite{6056691,9804195}, the power consumption of each BS in ON mode is modeled via a linear approximation approach. We denote by $P^{\text{ON}}_n$ the power consumed by BS $n$ to serve all the associated users, which can be expressed as follows:
\begin{align}
    P^{\text{ON}}_n = \sum_{m \in\mathcal{S}^{\text{BS}}_n}\kappa_{m,n}w_{m,n} + P^s_n \label{eq:P_on}
\end{align}
where $\kappa_{m,n}$ is the unit of radiated power needed by BS $n$ for serving user $m$ and $P^s_n$ is the static power consumption of BS $n$ (e.g., cooling). Let $P^{\text{OFF}}_n$ be the power consumption required by BS $n$ in OFF mode. Based on $a_n$, the total power consumption of BS $n$ denoted by $P^{\text{total}}_n$ can be written as follows:
\begin{align}
    P^{\text{total}}_n = a_nP^{\text{ON}}_n +(1-a_n)P^{\text{OFF}}_n. \label{eq:P_total}
\end{align}
Furthermore, we denote by $P^{\text{renew}}_n$ the power that can be harvested by BS $n$ from the renewable energy sources. To reduce the carbon emissions generated by non-renewable energy sources, we assume that each BS consumes renewable power as much as possible. Let $P^{\text{non-renew}}_n$ be the power consumption of non-renewables by BS $n$, which can be written as follows:
\begin{align}
    P^{\text{non-renew}}_n = \max (P^{\text{total}}_n - P^{\text{renew}}_n ,0).
\end{align}
In this paper, we focus on minimizing the utilized non-renewable power by the system in order to reduce its carbon footprint.
\begin{remark}
  The highly variable nature of renewable energy makes its availability varied considerably in time and space. Thus, it is not straightforward how a complicated heterogeneous network can efficiently utilize renewables when considering the temporal and spatial dimensions of renewable energy. 
\end{remark}

%% file: 3-MINLP.tex

Based on the system model defined in Section~\ref{sec:sys}, our goal is to design a scheme that chooses 1) $a_n$'s: the mode of each BS and 2) $w_{m,n}$'s: the association of users and BSs to minimize the non-renewable power consumption of all the base stations, which is given by $\sum _{n \in \mathcal{N}}P^{\text{non-renew}}_n$. In particular, the function $\sum _{n \in \mathcal{N}}P^{\text{non-renew}}_n$ is equal to
\begin{align*}
 \sum _{n \in \mathcal{N}}\max (a_nP^{\text{ON}}_n +(1-a_n)P^{\text{OFF}}_n- P^{\text{renew}}_n ,0),
\end{align*}
where $\sum _{n \in \mathcal{N}}\max (a_nP^{\text{ON}}_n +(1-a_n)P^{\text{OFF}}_n- P^{\text{renew}}_n ,0)$ is nonlinear due to the composition of the quadratic function (i.e., $a_nw_{m,n}$'s in $a_nP^{\text{ON}}_n$) and the max function.


Now, we introduce the joint user association and BS ON-OFF switching problem that minimizes the non-renewable power consumption of the network via the following mixed integer nonlinear programming (MINLP): 
\begin{align}
     \text{({\bf P1})} \ \min \ &\sum _{n \in \mathcal{N}}\max (a_nP^{\text{ON}}_n +(1-a_n)P^{\text{OFF}}_n- P^{\text{renew}}_n ,0) \nonumber\\
    s.t. \  & \sum_{n \in \mathcal{S}^{\text{user}}_m}w_{m,n} = 1,\ \forall m  \in \mathcal{M}; \label{eq:obj2}\\
     & \sum_{m \in \mathcal{S}^{\text{BS}}_n}w_{m,n} \leq B_n, \ \forall n  \in \mathcal{N}; \label{eq:obj3}\\
     & w_{m,n} \leq a_n, \ \forall m  \in \mathcal{M}, \ \forall n  \in \mathcal{N}; \label{eq:obj4}\\
     & w_{m,n} \in \{0,1\}, \ \forall m  \in \mathcal{M}, \ \forall n  \in \mathcal{N};\label{eq:obj5}\\
     & a_n \in \{0,1\}, \ \forall n  \in \mathcal{N}\backslash \{0\}; \label{eq:obj6} \\
     & a_0 = 1, \label{eq:a0}
\end{align} 
where $P^{\text{ON}}_n$ is a function of $w_{m,n}$'s defined in \eqref{eq:P_on}. Also, \eqref{eq:a0} ensures that MBS is always in ON mode. The formulated {\bf P1} problem is a MINLP which combines challenges of handling the nonlinearity of the objective function with combinatorial explosion of integer variables, which is in general challenging to solve~\cite{lee2011mixed}.


%% file: 4-MILP.tex
In this section, rather than directly minimizing the objective function of {\bf P1}, we use a derived upper bound to approximate {\bf P1}'s objective function $\sum _{n \in \mathcal{N}}P^{\text{non-renew}}_n$. We then propose a carbon-aware scheme by formulating a mixed integer linear programming (MILP) which efficiently minimizes the derived approximation of objective function in {\bf P1}. 

In the following, we present Lemma~\ref{lemma1} to bound $\sum _{n \in \mathcal{N}}P^{\text{non-renew}}_n$, where the proof of Lemma~\ref{lemma1} is provided in Appendix~\ref{proof_lemma1}.
\begin{lemma}\label{lemma1}
    For any two numbers $a,b \in \{0,1\}$, we have the following inequality:
    \begin{align}
        ab \leq \frac{a+b}{2}.
    \end{align}
\end{lemma}

By Lemma \ref{lemma1} and the fact that $a_n,w_{m,n} \in \{0,1\}$, we first bound the function $P^{\text{total}}_n$ as follows:
\begin{align}
    &P^{\text{total}}_n = \sum_{m \in\mathcal{S}^{\text{BS}}_n}\kappa_{m,n}a_nw_{m,n} + (P^s_n - P^{\text{OFF}}_n)a_n+P^{\text{OFF}}_n \label{eq:upper1}\\
    &\leq \sum_{m \in\mathcal{S}^{\text{BS}}_n}\kappa_{m,n}\cdot\frac{a_n+w_{m,n}}{2} + (P^s_n - P^{\text{OFF}}_n)a_n+P^{\text{OFF}}_n \label{eq:upper2}\\
    &= \sum_{m \in\mathcal{S}^{\text{BS}}_n}\frac{\kappa_{m,n}}{2}w_{m,n} + (\sum_{m \in\mathcal{S}^{\text{BS}}_n}\frac{\kappa_{m,n}}{2}+P^s_n - P^{\text{OFF}}_n)a_n+P^{\text{OFF}}_n \nonumber
\end{align}
where \eqref{eq:upper1} follows from \eqref{eq:P_on} and \eqref{eq:P_total}. We denote by $\tilde{P}^{\text{total}}_n$ the upper bound of $P^{\text{total}}_n$, i.e., $\tilde{P}^{\text{total}}_n = \sum_{m \in\mathcal{S}^{\text{BS}}_n}\frac{\kappa_{m,n}}{2}w_{m,n} + (\sum_{m \in\mathcal{S}^{\text{BS}}_n}\frac{\kappa_{m,n}}{2}+P^s_n - P^{\text{OFF}}_n)a_n+P^{\text{OFF}}_n$. Thus, the objective function of {\bf P1} can be bounded as follows:
\begin{align}
   \sum_{n \in \mathcal{N}}P^{\text{non-renew}}_n & = \max (P^{\text{total}}_n - P^{\text{renew}}_n ,0) \nonumber\\
   & \leq  \sum_{n \in \mathcal{N}}\max (\tilde{P}^{\text{total}}_n - P^{\text{renew}}_n, 0) \label{eq:upper4}.
\end{align}

The optimization problem for minimizing the approximation of objective function $\sum_{n \in \mathcal{N}}P^{\text{non-renew}}_n$ can be presented as follows:
\begin{align}
   (\textbf{P2}) \ \min \ & \sum_{n \in \mathcal{N}}\max (\tilde{P}^{\text{total}}_n - P^{\text{renew}}_n, 0) \label{op1}\\
    s.t. \ & \text{\eqref{eq:obj2} to \eqref{eq:a0}}.
\end{align} 
Similar to {\bf P1}, the MINLP formulation of {\bf P2} is also hard to solve in general. 

To efficiently obtain the optimal solution for {\bf P2}, we transform its objective function into a linear function by introducing a dummy variable denoted $y_n$ for each $n \in \mathcal{N}$ with some additional constraints. More specifically, each introduced variable $y_n$ is lower-bounded by $\max (\tilde{P}^{\text{total}}_n - P^{\text{renew}}_n, 0)$ and the newly formulated problem minimizes the objective function given by $\sum_{n \in \mathcal{N}}y_n$. Our proposed carbon-aware scheme aims at choosing $a_n$'s and $w_{n,m}$'s by solving this transformed optimization problem, which can be presented by a MILP as follows:
\begin{align}
    (\textbf{P3}) \ \min \ & \sum_{n \in \mathcal{N}}y_n\\
     s.t. \  & y_n \geq \tilde{P}^{\text{total}}_n - P^{\text{renew}}_n, \  \forall n \in \mathcal{N}; \label{eq:P3_1}\\
            & y_n \geq 0, \ \forall n \in \mathcal{N}; \label{eq:P3_2}
\end{align}
and subject to \eqref{eq:obj2} to \eqref{eq:a0}, in which \eqref{eq:P3_1} and \eqref{eq:P3_2} are due to the additional constraints that ensure $y_n \geq \max (\tilde{P}^{\text{total}}_n - P^{\text{renew}}_n, 0) $ for $\forall n \in \mathcal{N}$.

The following theorem, whose proof is provided in Appendix~\ref{proof_thm}, shows the equivalence of \textbf{P2} and \textbf{P3}.
\begin{theorem}[Equivalence]\label{thm}
    The optimization problems \textbf{P2} and \textbf{P3} are equivalent, i.e., a feasible solution of one problem can be used to construct a feasible solution of the other problem. Moreover, the optimal values of \textbf {P2} and \textbf {P3} are identical.
\end{theorem}

If the collection of $w_{m,n}^{*}$'s, $a^{*}_n$'s and $y^{*}_n$'s is the optimal solution of {\bf P3}; by Theorem \ref{thm}, then our proposed scheme obtained by solving {\bf P2} is the collection of $w_{m,n}^{*}$'s and $a^{*}_n$'s. Unlike the MINLP (e.g., {\bf P1} and {\bf P2}), the MILP formulation of {\bf P3} in our proposed scheme can be solved efficiently by robust commercial solvers, e.g., CPLEX~\cite{nickel2022ibm}, Gurobi~\cite{gurobi} and Matlab Optimization Toolbox~\cite{MatlabOTB}. Therefore, the MILP formulation of {\bf P3} via the linearization of objective function allows us to obtain the optimal solution of {\bf P2} efficiently.

%% file: 5-analysis.tex
In this section, we demonstrate the impact of the proposed carbon-aware scheme by simulation studies. We evaluate the effectiveness of the proposed scheme in terms of the non-renewable power. We consider a cellular network composed of a MBS and $N = 8$ SBSs. BS $0$ has the coverage radius of $r_0 = 600m$, and each BS $n \in \mathcal{N}\backslash\{0\}$ has the coverage radius of $r_n = 200m$. In the unit of $m$, BS $0$ (i.e., MBS) is located at $(0,0)$; and each BS $n \in \mathcal{N}\backslash\{0\}$ (i.e., SBSs) is located at $(200,200)$, $(-200,-200)$, $(200,-200)$, $(-200,200)$, $(0,-400)$, $(0,400)$, $(400,0)$, $(-400,0)$ respectively. 

There are in total $M = 300$ mobile users in the coverage area of MBS, where the coordinate of each user $m$ is randomly generated in an uniform manner. BS $0$ can serve up to $B_0 = 200$ users; and each BS $n \in \mathcal{N}\backslash\{0\}$ can serve up to $B_n = 60$ users. Each BS has the static power consumption $P^s_n = 2000$W; and the power consumption in OFF mode $P^{\text{OFF}}_n = 0$W. Based on~\cite{9804195}, each BS has $\kappa_{m,n} = 18d_{m,n}^{2.6}$(W) where $d_{m,n}$ is measured in the unit of $km$.   

\textbf{Wind Power Generation Model:} We assume that each BS is powered by a small wind turbine for the source of renewable power. We denote by $\ell_n$ the radius of wind turbine at BS $n$. Then, the cross-sectional area swept by the wind turbine $A_n$ at BS $n$ is given by $A_n = \pi \ell_n^2$. The wind speed denoted by $v_n$ at each BS $n$ is modeled via Weibull distribution, which can well describe the statistical properties of the wind speed~\cite{MAHMOOD202079}. Given wind speed $v_n$ at each BS $n$, the wind power that can be harvested by the BS is given by the formula $P^{\text{renew}}_n = \frac{1}{2}\rho A_n v_n^3$, where $\rho$ is the air density.

\begin{figure}[t]
    \centering
    \includegraphics[width = \linewidth]{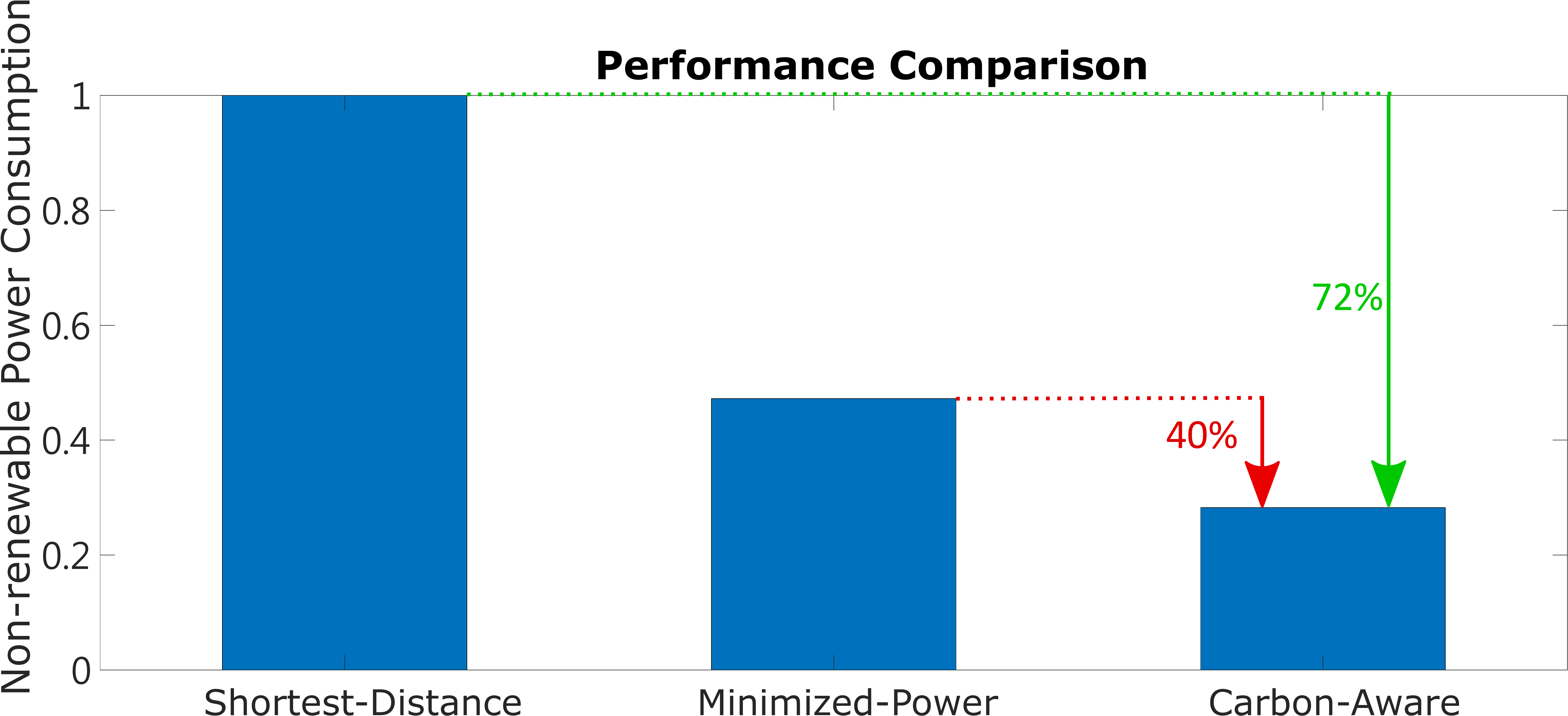}
    \caption{Numerical evaluations for the normalized non-renewable power where each BS $n$ is powered by the wind turbine with the radius of $\ell_n =1.5m$.}  
    \label{fig:r15}
\end{figure}

\begin{figure}[t]
    \centering
    \includegraphics[width = \linewidth]{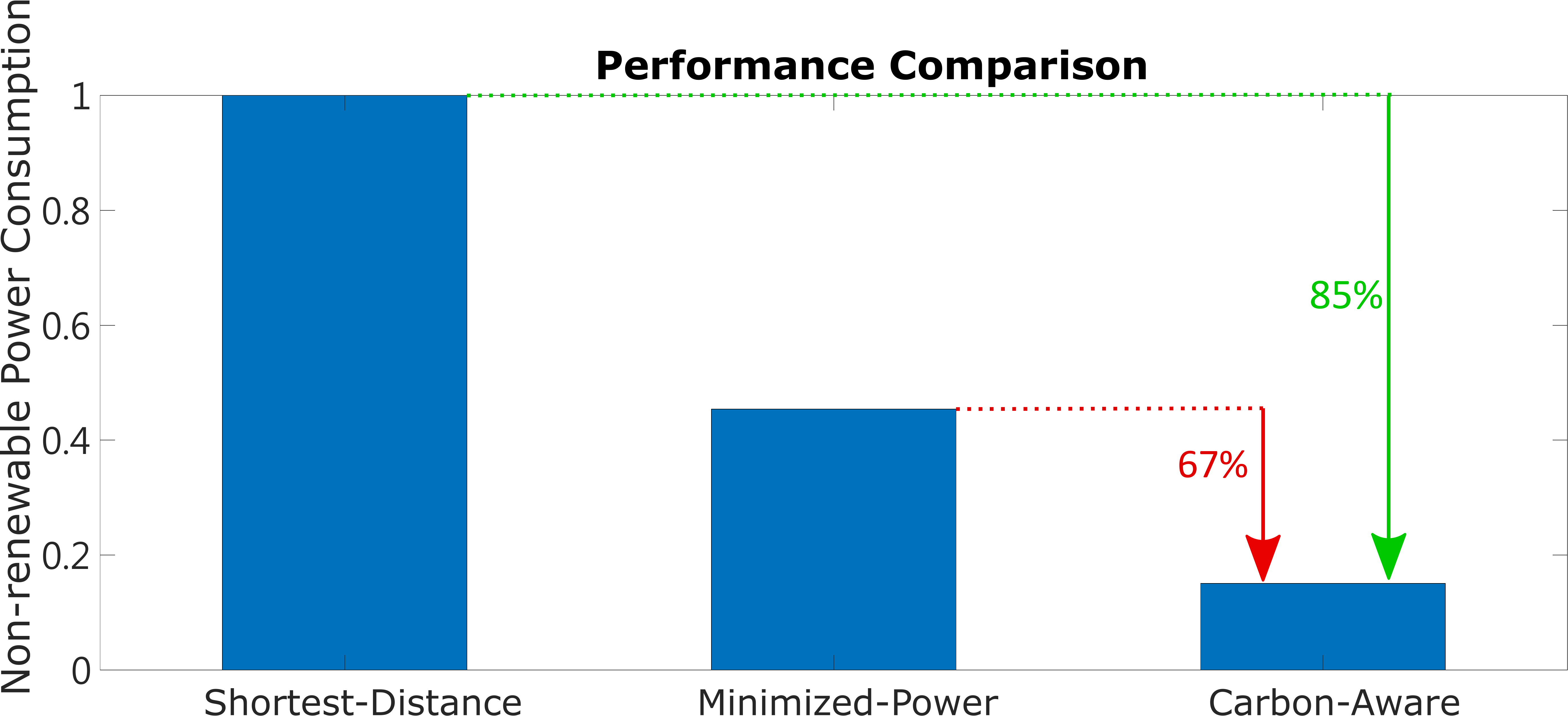}
    \caption{Numerical evaluations for the normalized non-renewable power where each BS $n$ is powered by the wind turbine with the radius of $\ell_n =3m$.}  
    \label{fig:r30}
\end{figure}

\begin{figure}[t]
    \centering
    \includegraphics[width = \linewidth]{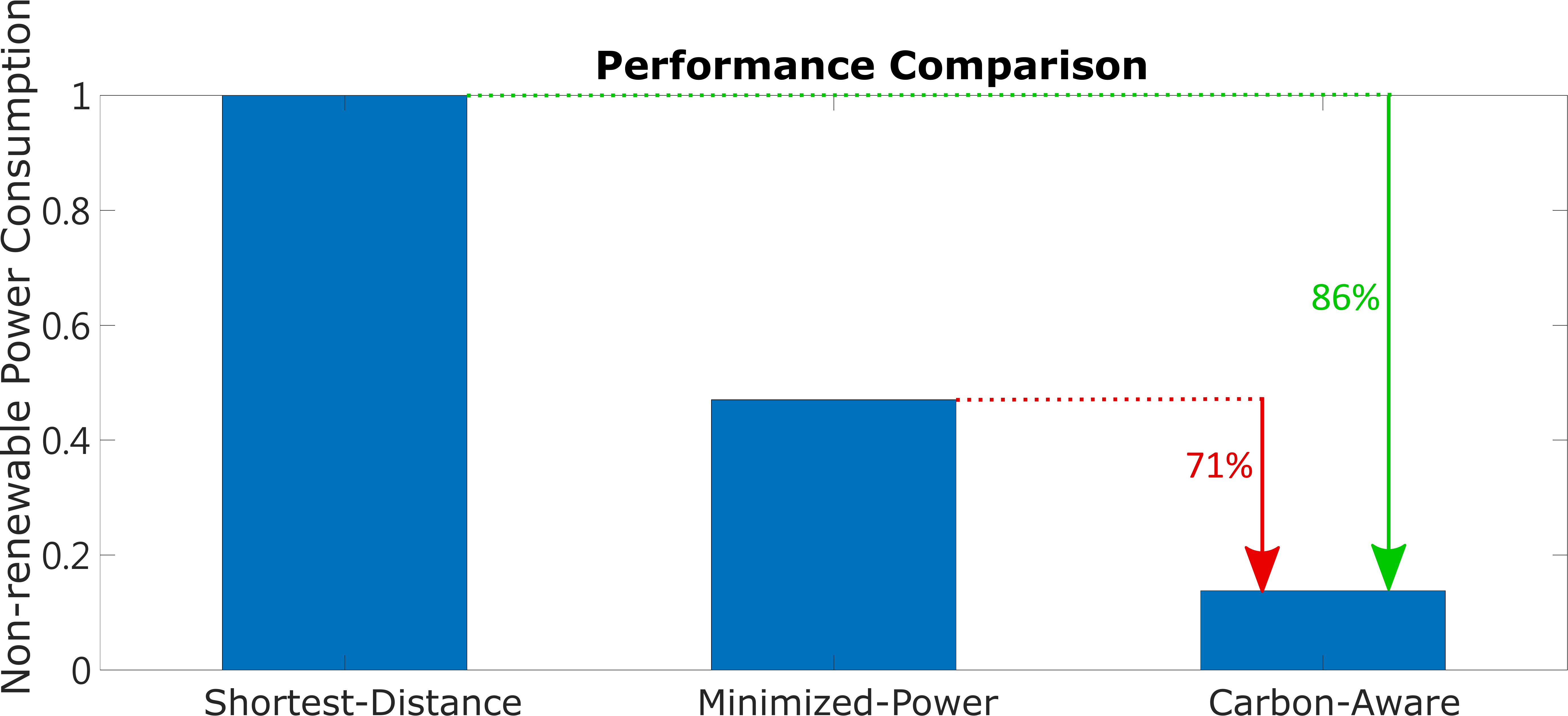}
    \caption{Numerical evaluations for the normalized non-renewable power where each BS $n$ is powered by the wind turbine with the radius of $\ell_n =4.5m$.}  
    \label{fig:r45}
\end{figure}

In this analysis, we assume that the wind speed at each BS is generated by Weibull distribution with the shape parameter of $2.081$ and the scale parameter of $6.69$ based on the data in~\cite{MAHMOOD202079}. The density of air in the system is $\rho = 1.225kg/ m^3$. The resulting MILP problems in our proposed scheme and the baselines are solved by \texttt{intlinprog} from Matlab Optimization Toolbox~\cite{MatlabOTB}. The following baselines are considered to compare with the proposed carbon-aware scheme:
\begin{enumerate}[leftmargin =*]
    \item {\bf Shortest-Distance Scheme:} Each BS $n$ is always switched to ON mode, i.e., $a_n =1, \forall n \in \mathcal{N}$. Select $w_{m,n}$'s such that each UE $m$ is served by its closest BS to ensure the strength of received signal power.
    \item {\bf Minimized-Power Scheme:} Select $a_n$'s and $w_{m,n}$'s that minimize the total power (renewables + non-renewables) utilized by all the BSs. This selection can be obtained by formulating a MILP which minimizes $\sum_{n \in \mathcal{N}}P^{\text{total}}_n$ subject to \eqref{eq:obj2} to \eqref{eq:a0}.  
\end{enumerate}
In terms of the normalized non-renewable power consumption averaged over $500$ simulations, Fig.~\ref{fig:r15}, Fig.~\ref{fig:r30} and Fig.~\ref{fig:r45} provide the performance comparison of the carbon-aware scheme with the baselines for the radius $\ell_n = 1.5m, 3m$ and $4.5m$ respectively.\footnote{The power consumption will scale up in the size of the system. Hence, we only focus on the normalized non-renewable power consumption in this analysis.} Then, we conclude the followings:
\begin{itemize}[leftmargin =*]
    \item The proposed carbon-aware scheme provides a significant  non-renewable power consumption reduction up to $86\%$ and $71\%$ over the shortest-distance scheme and the minimized-power scheme respectively.
    \item As $\ell_n$ increases (i.e., more available wind power), the proposed carbon-aware scheme reduces more non-renewable power consumption compared with two baselines, which demonstrates the effectiveness of carbon-aware scheme in utilizing renewable power.
    \item The huge improvement over the minimized-power scheme indicates the importance toward the design of carbon-oriented technology beyond the traditional approach (i.e., energy efficiency) to meet the goal of carbon neutrality.
\end{itemize}
\begin{remark}
We would like to emphasize that our proposed scheme can potentially yield a huge saving of money by just reducing the emissions of carbon dioxide. For example, about $2$ million 5G base stations have been deployed in China~\cite{china5G}. Without BS ON-OFF switching, if each BS in China consumes the power of $3000$W in $24$ hours every day, then they result in total energy consumption of $5.3\times 10^{10}$kW$\cdot$h. Assuming that carbon intensity of grid energy in China is $500$g per kW$\cdot$h and renewables account for $20\%$ of energy consumption, BSs would contribute $2.1\times 10^7$ tons of carbon emissions. Since each ton of carbon emissions cost the tax of $9$ USD in China according to The World Bank~\cite{worldbank} , there can be a money saving of $2$ billion USD annually by using our scheme.
\end{remark}

%% file: 6-conclusion.tex
In this paper, in order to reverse the effects of global warming, we consider a problem of joint user association and base station ON-OFF switching with the objective of minimizing non-renewable power consumption. Through a linear approximation, we proposed an efficient carbon-aware scheme for heterogeneous cellular networks, which utilizes the availability information of renewable power at each base station to effectively reduce carbon footprint of networks. The numerical analysis in our paper demonstrates that the proposed scheme can effectively reduce the non-renewable power consumption by $86\%$ and save billions of US dollars. This work motivates us to take the carbon-related information into account when designing the next-generation cellular network toward the objective of carbon neutrality by 2050 (e.g.,~\cite{yang2022carbon,10003083}). Beyond this work, another future research direction is to consider other deployments of a cellular network, e.g., a distributed network of radio units, distributed units and centralized units that covers radio access network and core network. Designing a scheme to minimize the overall carbon emissions of whole system could be an interesting and challenging problem.

%% file: Appendix.tex
\section{Proof of Lemma \ref{lemma1}}\label{proof_lemma1}
Since $a$ and $b$ are non-negative, the AM-GM inequality~\cite{Hoffman1981} holds as $\frac{a^2+b^2}{2} \geq \sqrt{a^2b^2} = ab$. With the fact of $a,b \in \{0,1\}$, we have $\frac{a+b}{2} = \frac{a^2+b^2}{2}\geq ab$ which concludes the proof.
\section{Proof of Theorem \ref{thm}}\label{proof_thm}
First, we assume that a collection of $w_{m,n}$'s and $a_n$'s is a feasible solution of \textbf{P2}. Then, we define $y_n$ as follows: 
    \begin{align}
        y_n = \sum_{n \in \mathcal{N}}\max (\tilde{P}^{\text{total}}_n - P^{\text{renew}}_n, 0).
    \end{align} 
    Each of the constraints defined in \eqref{eq:P3_1} and \eqref{eq:P3_2} are satisfied due to the definition of max function. With the fact that $w_{m,n}$'s and $a_n$'s satisfy the constraints defined in \eqref{eq:obj2} to \eqref{eq:a0}, the collection of $w_{m,n}$'s, $a_{n}$'s and $y_n$'s is a feasible solution of \textbf{P3}. The value of objective function in \textbf{P2} is $\sum_{n \in \mathcal{N}}\max (\tilde{P}^{\text{total}}_n - P^{\text{renew}}_n, 0)=\sum_{n \in \mathcal{N}}y_n$, which is the same as the value of objective function in \textbf{P3}. It follows that the optimal value of \textbf{P2} is greater than or equal to the optimal value of \textbf{P3}.
   
    Conversely, we assume that the collection of $w_{m,n}$'s, $a_{n}$'s and $y_n$'s is a feasible solution of \textbf{P3}. It is clear that the collection of $w_{m,n}$'s, $a_{n}$'s is a feasbile solution of \textbf{P2}. Because the constraints defined in \eqref{eq:P3_1} and \eqref{eq:P3_2} are satisfied, we have $y_n \geq \max (\tilde{P}^{\text{total}}_n - P^{\text{renew}}_n, 0)$ for each $n$. Therefore, the value of objecitve function  $\sum_{n\in \mathcal{N}}y_n$ in \textbf{P3} is greater than or equal to the value of objective function in \textbf{P2}, which is given by $\sum_{n \in \mathcal{N}}\max (\tilde{P}^{\text{total}}_n - P^{\text{renew}}_n, 0)$. It follows that the optimal value of \textbf{P3} is greater than or equal to the optimal value of \textbf{P2}, which completes the proof.